\documentclass{article}
\usepackage{moreverb}
\usepackage{tikz}
\usepackage[numbers]{natbib}
\usepackage{booktabs}
\usepackage{amsmath}
\usetikzlibrary{arrows.meta}
\usepackage{graphicx} 
\usetikzlibrary{shapes,positioning}
\begin{document}

\title{Predicting subgroup treatment effects for a new study: Motivations, results and learnings from running a data challenge in a pharmaceutical corporation}

\date{}
\author{Björn Bornkamp\thanks{Global Drug Development, Novartis Pharma AG, Basel, Switzerland}
\and Silvia Zaoli\footnotemark[1]
\and Michela Azzarito\footnotemark[1]
\and Ruvie Martin\thanks{Global Drug Development, Novartis Pharmaceuticals Corporation, New Jersey, USA}
\and Carsten Philipp Müller\thanks{Data \& Digital, Beyond Conception GmbH, Altendorf, Switzerland}
\and Conor Moloney\thanks{Global Drug Development, Novartis Pharma AG, Dublin, Ireland}
\and Giulia Capestro\footnotemark[1]
\and David Ohlssen\footnotemark[2]
\and Mark Baillie\footnotemark[1]}







\maketitle

\abstract{
We present the motivation, experience and learnings from a data challenge conducted at a large pharmaceutical corporation on the topic of subgroup identification.  The data challenge aimed at exploring approaches to subgroup identification for future clinical trials. To mimic a realistic setting, participants had access to 4 Phase III clinical trials to derive a subgroup and predict its treatment effect on a future study not accessible to challenge participants. 30 teams registered for the challenge with around 100 participants, primarily from Biostatistics organisation.  We outline the motivation for running the challenge, the challenge rules and logistics. Finally, we present the results of the challenge, the participant feedback as well as the learnings, and how these learnings can be translated into statistical practice.
}


\section{Introduction}
\label{sec:intro}

In many computational disciplines data challenges are a central paradigm; holding regular competitions has led to substantial improvements over time on the underlying predictive modelling task. For example the ImageNet challenges \citep{deng2009imagenet} for image processing, DREAM challenges \citep{dream:Stolovitzky:2016} in biomedicine, the CASP challenges \citep{Kryshtafovych:CASP:2019} for protein structure prediction or the TREC challenge \citep{Voorhees:Trec:2005} for information retrieval. Indeed, many aspects of today's life are in one way or another influenced by the result of such data challenges. Donoho (2017)\cite{Donoho2017-tw} describes data challenges and the underlying common task framework (CTF) \citep{liberman2010} as the "secret sauce" for the success of predictive modelling in many areas over the past decades. The CTF has three key elements: (1) a common task, (2) shared data, and (3) a standard evaluation. The common task with shared data/information and a standardized evaluation enables discussion and learning on the problem within teams and, after closing the challenge, also across teams, based on a common ground. The framework includes a competitive element, which is motivating, but also contains a community element for the participants, who all work on the same common problem. 

Another interesting aspect is that it allows assessment of the breadth of solutions across teams. For exploratory analyses in drug development projects, one usually only sees one solution proposed by a single person (or team) working on the data. It would often be of interest to assess the variability of conclusions across multiple teams, to assess reliability. 

In addition, from the perspective of a larger organization, data challenges are a way of training and learning, \textit{i.e.}, allowing to improve the skill-set of the overall population of participants on the particular problem, but also learning for the organization as a whole. For participants the learning is quite effective as it is "learning by doing", and at the same time it happens in the "safe space" of a data challenge.  

Motivated by this set of benefits, we decided to conduct a data challenge on the topic of subgroup identification in randomized clinical trials across the quantitative scientists within a pharmaceutical corporation. Within a drug development setting, subgroup identification based on clinical trial data is seen as one of the most difficult problems to get right, and has been described as "the hardest problem there is" \cite{ruberg2021assessing}. It is well-known to be prone to conclusions based on statistical analyses that later could not be reproduced in follow-up trials \cite{Yusuf1991, assm:poco:enos:2000, slei:2000, rothwell2005subgroup, aust:2006, laga:2006, alosh:2017, ema:2019}. The reasons for this are multi-fold. One of them is insufficient sample size: Clinical trials are typically designed to make statistically reliable statements about the overall effect but not about subgroup treatment effects. Another issue is multiplicity/selective inference: Subgroup treatment effects are subject to variability, so that searching across many subgroups will naturally lead to identification of subgroups with random-low or random-high results. This problem increases in severity with the number of subgroups that are investigated. Both previous difficulties cannot primarily be solved by utilizing a particular data analytic method alone, it depends also on how the analysis is being planned, conducted and its results being interpreted. This is particularly important in subgroup identification or subgroup analyses, as those are typically not pre-defined but performed post-hoc. This induces flexibility and leads to many choices for the statistical analysis, which the analyst may decide upon based on personal biases, theory or interpretations/assumptions \cite{Silberzahn2018}, potentially increasing the variability of results and decreasing the quality. Paradoxically, exploratory investigations hence require increased self-discipline and rigor. In addition, few standards have emerged on how to approach subgroup identification or exploration of treatment effect heterogeneity, with many novel methods still being published and proposed. This further adds to a large number of choices from the perspective of the analyst.

Despite these challenges with subgroup identification or exploration of treatment effect heterogeneity, there is a natural desire and scientific mandate to explore subgroup treatment effects for well-informed clinical decision making, which puts the analyst in a challenging situation. Better understanding treatment effect variation may also help project teams understand the mode of action of the drug and the underlying disease indication better. The importance of exploring treatment effect variation is also documented in the large volume of recent papers and guidance on the topic \cite{lipk:dmit:agos:2017,ema:2019,kent2020:path,nie2021quasi}. 

The outline of this paper is as follows. In Section \ref{sec:rulebook} we will describe the challenge task, its rules, as well as the provided data. In Section \ref{sec:results} we will review and discuss the results of the challenge, including also participant feedback. We conclude with a summary of the key learnings, and how these learnings could be translated into statistical practice.

\section{The subgroup identification data challenge}
\label{sec:rulebook}

\subsection{Challenge Task} 

Instead of running a simulation based challenge, where the truth is set up by the organizers, we wanted to emulate a realistic subgroup identification scenario that project teams may face, where the "truth" is not known. Hence the challenge case was selected by searching for an internal drug program, where some clinical trials are already available, but where there is also another clinical trial still concurrently ongoing in the same indication for the same drug that can be used for scoring. This allows for making real "prospective" predictions by the participants and also allows for a real "hold-out" evaluation of the solutions predicted by the participants. In many ways this reflects the real situation of deriving a subgroup with promising treatment effect based on existing data and then assessing it in a follow-up study. In addition we searched for a program where a number of previous trials were available so that there was a chance to identify a subgroup with an increased effect. 

For this purpose Cosentyx (secukinumab, AIN457) in the psoriatic arthritis (PsA) indication was identified as a suitable case \citep{blair2021secukinumab}. Cosentyx 
is a human monoclonal interleukin (IL)-17A antagonist indicated for the treatment of PsA, as well as moderate to severe plaque psoriasis, ankylosing spondylitis, non-radiographic axial spondyloarthritis, as well as two juvenile idiopathic arthritis subtypes. PsA is a chronic inflammatory disease that affects peripheral and axial joints, entheses, and the skin, and is often associated with impaired physical function and poor quality of life. Symptoms of inflammatory arthritis include stiffness, pain, and swelling of the joints, restricted motions, and reduced physical strength. The main endpoint used in PsA is the American College of Rheumatology (ACR) score \cite{felson1993american}, which is a composite endpoint, combining measurements of number of tender joints, number of swollen joints, patients assessment of PsA pain, patient's and physician's global assessment of disease activity as well as patient's assessment of physical function. Primary endpoints in clinical trials are often ACR20 or ACR50, which are binary endpoints indicating an improvement of at least 20 or 50\% in tender and swollen joint count and of at least 20 or 50\% in at least 3 of the 5 other ACR components versus baseline.

Participants gained access to the anonymized data of four randomized, double-blind, placebo-controlled clinical trials for Cosentyx in PsA (see Table \ref{tab:tabone}). The task of the participants was to work on these data and to define a subgroup with increased treatment effect that would be assessed in the INVIGORATE 2 trial whose access was still restricted.  Following the primary outcome specified in INVIGORATE 2,  ACR50 at week 16 was used to assess the treatment effect in the identified subgroup. In addition to the subgroup definition, we also asked the participants to provide predictions of the treatment effect $\delta_{pred}$ for the proposed subgroup on the risk difference scale, as well as an estimate of the standard deviation of their prediction, $\sigma_{pred}$. Both quantities together are important as they are needed for planning "the next trial": They allow to evaluate the probability of success or assurance \citep{o2005assurance} of a new trial to be planned.

\subsection{Shared Data}
\label{sec:data}

\subsubsection{Clinical trial data}
Details on the utilized clinical trials can be seen in Table \ref{tab:tabone}. The anonymized data for the FUTURE 2-5 studies were made available to participants on the company internal data science platform. Data from the new trial, the INVIGORATE 2 study, once available, were only accessible on the restricted regulatory data platform, were participants had no access. 

\begin{table}
    \caption{Summary of clinical studies included in the data challenge, including details on treatment regimes within each trial. The abbreviations NL and L stand for no-load and loading regimen and s.c. and i.v. stand for subcutaneous and intravenous (for details see main text). \label{tab:tabone}}
    \centering
        \begin{tabular}{rrrr} 
        \toprule
         Study & Treatments & Regimen & sample size \\ 
         \midrule
         FUTURE 2 (NCT01752634) & Placebo (PBO), 75mg, 150mg, 300mg & s.c., L & $\sim$100 per arm \\ 
         FUTURE 3 (NCT01989468) & PBO, 150mg, 300mg & s.c., L & $\sim$135 per arm \\
             FUTURE 4 (NCT02294227) & PBO, 150mg (L), 150mg (NL) & s.c., L \& NL & $\sim$106 per arm \\
         FUTURE 5 (NCT02404350) & PBO, 150mg (L), 150mg (NL), 300mg (L) & s.c., L \& NL & $\sim$990 in 3:2:2:2 ratio \\
         INVIGORATE 2 (NCT04209205) & PBO, 6mg/kg + 3mg/kg & i.v. & $\sim$190 per arm \\ 
         \bottomrule
        \end{tabular}
\end{table}

The inclusion and exclusion criteria for all 5 studies were identical and the observed baseline characteristics very similar. A key difference was how Cosentyx was administered across the studies, see Table \ref{tab:tabone}. The standard regimen is a "loading" regimen (abbreviated by "L" in Table \ref{tab:tabone}); this consists of injections at baseline, week 1, 2, 3, 4 and then every four weeks thereafter. "NL" stands for a no-loading regimen, which consists of administrations every four weeks starting from baseline. The FUTURE 2-5 studies used subcutaneous (s.c.) injection, while in the INVIGORATE 2 study intravenous (i.v.) infusion (6mg/kg at baseline, then 3mg/kg every four weeks thereafter) is used. Based on the existing clinical experience and understanding of Cosentyx pharmacokinetics, this regimen was expected to provide a similar pharmacokinetic profile and clinical response as the 300mg loading (L) regimen. An explanation and justification of the i.v. regimen and its similarity to the s.c. regimen was part of the protocol of the INVIGORATE 2 trial, which was provided to the participants.

Participants were provided with the data in CDISC ADaM data standard \cite{cdisc:adam}. The data consisted of 2,148 patients across four studies - FUTURE 2, 3, 4 and 5. The data was split across three standard files in the ADaM format - ADSL, ADBS and ADEFF. ADSL contained 28 baseline covariates. These covariates provided some basic information like AGE, SEX, BMI; information on medical history such as time since first PsA diagnosis, presence of psoriasis, presence of polyarticular arthritis; covariates relating to other treatments and medications like corticosteriod use, TNF alpha inhibitors. ADBS contained 71 baseline covariates relating to lab values, clinical measurements, quality of life measurements and efficacy measurements. ADEFF contained 21 efficacy endpoints, including ACR50, across 7 time points (weeks 1, 2, 3, 4, 8, 12, 16), where available. These endpoints included ACR component variables such as HAQ disability index, CRP, phsycian's global assessment of disease activity, subjects global assessment of disease activity, tender joint total score for PsA 78 joints and swollen joint total score for PsA 76 joints. 

A template Git repository including scripts to access the datasets and instructions on the submission procedure were provided to the participants. Teams were expected to work in this repository and provide their submission by filling out information in pre-defined places in a submission document, which was provided as part of the template repository. 

\subsubsection{Further information sources}

Participants obtained the protocols for all involved studies, data-set specifications and background material on subgroup identification (\textit{i.e.} literature references and example code). Two clinical experts were identified to help with potential questions that participants may have on the compound, disease or subgroup relevance more generally. Participants could raise their questions either via chat or direct contact.

\subsection{Standard Evaluation}

\subsubsection{Submission of results}
The submission file in the Git repository required teams to enter
\begin{itemize}
    \item[(a)] the subgroup definition (max 100 characters using a valid R logical expression, e.g., “CRP $> 2.5\  \&$ AGE $< 40$” or more complex subgroups defined by scores such as "$0.1$*AGE $- 0.5$*(MTXUSE == 'Yes')) $> 4.7$"
    \item[(b)] their values of $\delta_{pred}$ and $\sigma_{pred},$ and
    \item[(c)] a description of the methodology used to derive the subgroup, including clinical/biological justification (max 300 words).
\end{itemize}

A proposed subgroup was considered valid only if it included at least 10 subjects on treatment and 10 on control in each group (subgroup and complement), on the INVIGORATE 2 trial. As the participants had no access to the baseline covariates of the INVIGORATE 2 trial, two test submissions were performed before the final submission, where the size of the subgroups proposed by the teams was assessed on the INVIGORATE 2 trial and reported back.

\subsubsection{Scoring rules}

We scored the submission based on a quantitative and qualitative score. The quantitative score aimed at measuring the quality of the proposed subgroup and of the treatment effect prediction; whereas the qualitative score aimed at rewarding ``good data science practice"~\citep{Baillie:2022}.

The quantitative score was used to rank all teams, and the top 3 were awarded a prize. An additional prize was given to best team out of the top 5, according to the qualitative score.

The quantitative score $Z$ was computed as the mean of two scores. The first score evaluated the quality of the proposed subgroup in the INVIGORATE 2 trial. How to best assess this is not obvious. Naively one may just want to use the magnitude of the estimated treatment effect of the proposed subgroup in the INVIGORATE 2 trial. This approach, however, would be problematic as teams may deliberately choose small subgroups, which, by the larger variability, may have very large treatment effects (or very small treatment effects), making the challenge results prone to chance. For this reason we rather assessed the evidence for treatment effect modification in the subgroup, by using the test statistic for the subgroup-by-treatment interaction. 

To compute this score, we fitted the following logistic model to the ACR50 response at week 16 in INVIGORATE 2 patients: 
\begin{equation}
    logit(p_{i}) = \beta_{0} + \beta_{trt}t_{i} + \beta_{s}s_{i} + \beta_{interaction}t_{i}s_{i} + x_{i}^{'}\beta  
\end{equation} 
where $p_i$ is the probability of response of patient $i$, $t_i$ is a binary variable indicating the arm  (0 – Placebo, 1 – Cosentyx ), $s_{i}$ is a binary variable indicating the subgroup membership (1 – subgroup, 0 – complement), and $x_{i}$ are two covariates also included in the primary analysis model for the INVIGORATE 2 trial (weight as a continuous covariate and whether the patient was naive to TNF-$\alpha$ inhibitors as binary covariate). 
 
According to this model, the treatment effect in subgroup (on logit scale) is $\beta_{trt} + \beta_{interaction}$, while the treatment effect in complement group is $\beta_{trt}$. Therefore, $\beta_{interaction}$ represents the modification of the treatment effect in the subgroup.  A value of $\hat{\beta}_{interaction} > 0$ implies that the treatment effect in the subgroup is larger than in the complement group. Therefore, we define our first score as
\begin{equation}
S_1 = \frac{\hat{\beta}_{interaction}}{s.e.(\hat{\beta}_{interaction})},
\end{equation}
where we divide by the standard error, to avoid that results are overly driven by chance.
To mitigate issues around complete separation in the model fitting (beyond the minimum size requirements on the subgroup) the logistic regression model was fitted using Firth's method \citep{firth1993bias}. In this analysis the primary intercurrent event, treatment (and study) discontinuation was handled in the same way as in the primary study analysis by a composite strategy, \textit{i.e.} patients who discontinue the treatment are considered as ACR50 non-responders \citep{ich:2019}.

The second score evaluated the prediction of the treatment effect ($\delta_{pred}$, $\sigma_{pred}$), given the observed value of the treatment effect on the risk difference scale in the subgroup, $\hat{\delta}$. The treatment effect estimate was based on applying standardization (to transform results to the risk difference scale) using the logistic regression model above \citep{ge2011covariate}. The score then assesses the log-likelihood of a Gaussian distribution centered at $\delta_{pred}$ with standard deviation $\sigma_{pred}$, evaluated at $\hat{\delta}$:
\begin{equation}
S_2 = -0.5\log(2\pi) - \log(\sigma_{pred}) - 0.5\frac{( \hat{\delta} - \delta_{pred} )^2}{\sigma^2_{pred}}   
\end{equation}

The two scores $S_1$ and $S_2$ were then standardized by robust estimators of location and scale across teams, so that they are on a comparable range of values. The score for team $j$ is given by
\begin{align}
    Z_{1,j} & = \frac{S_{1,j} -  \mathrm{median}(S_{1,1},...,S_{1,k})}{\mathrm{1.4826 MAD}(S_{1,1},...,S_{1,k})}\\
    Z_{2,j} & = \frac{S_{2,j} -  \mathrm{median}(S_{2,1},...,S_{2,k})}{\mathrm{1.4826 MAD}(S_{2,1},...,S_{2,k})}
\end{align}
Here $S_{1,j}$ and $S_{2,j}$ are the scores for team $j$ and $k$ is the total number of teams. MAD stands for median absolute deviation. The standard scaling factor $1.4826$ is used for the MAD (this ensures a consistent estimate of the standard deviation for Gaussian data). The reason for using robust estimators is that we anticipated potential extreme results, which then could influence the weighting of the two components.

The final score used to rank team $j$ is obtained by averaging the two scores:
\begin{equation}
 Z_j=\frac{Z_{1,j}+Z_{2,j}}{2}
\end{equation}

The R function that was used for scoring, \textit{i.e.}, to calculate $S_1$ and $S_2$, was provided to participants for transparency so that they could potentially use it as part of the model training.

\subsubsection{Qualitative evaluation: peer review}
 
The qualitative scoring involved a peer review of all team repositories based on good practices. Two criteria were defined to structure the review: (i) science and methodology - teams submitted a short methodological statement for review to help understand rationale and interpretation; (ii) project documentation and materials - the motivation of this criteria was to assess reproducibility and knowledge transfer. The latter criteria includes specific elements such as the clarity of the code (e.g. use of good coding practices and comments, informative README file, version control, etc.) and reproducibility (e.g. if a new user can easily and independently implement the approach from the documentation that is provided via the project repository and get the same results). 

The first criterion was assessed based on blinded review of the submitted methodological statement (point (c) above) by 3 senior associates from the biostatistics and clinical line function who were not involved with the data challenge. The second criterion was assessed by the organizing committee. The qualitative review was performed only for the top 5 teams according to the quantitative score.

\section{Challenge Results}
\label{sec:results}

98 participants registered for the challenge across 30 teams. 21 out of these 30 teams had at least one member working in the biostatistics line-function, but there were also teams from functions such as real world evidence, data science and artificial intelligence, as well as business analytics. The challenge started on the 13th December 2021 with a kick-off event, where the challenge and its rules were introduced. During the challenge there was one further event to explain details on the submission procedure. In addition two test submissions were done in order to get familiar with the submission procedure and to provide participants feedback regarding the size of their subgroup in the INVIGORATE 2 trial. The final submission was done on the 24th January 2022. After the winner announcement there were two sessions, where the winning teams presented their solutions and a panel discussion including organizers and participants was performed to discuss the results and learnings from the challenge.

\begin{figure}
\begin{center}
\begin{tikzpicture}[font=\small,thick]
 \node[draw,
    minimum height=0.6cm] (block1) { \tiny Teams registered for challenge ($n=30$)};
 \node[draw,
    below=of block1,
    minimum height=0.6cm
] (block2) { \tiny Incomplete submissions ($n=4$) };
 \node[draw,
    below left=of block1,
    minimum height=0.6cm
] (block3) { \tiny Complete submissions ($n=22$)};
\node[draw,
    below right=of block1,
    minimum height=0.6cm
] (block4) { \tiny No submission ($n=4$)};
\draw[rounded corners, thick]
  (block1) -- (block2)
  (block1) -- (block3)
  (block1) -- (block4);
\end{tikzpicture}
\end{center}
\caption{Summary of the participation to the challenge. Teams that did not modify the submission file in the Git repository were counted as "No submission". Among the 4 with incomplete submission files, none had a subgroup justification, all had a subgroup definition, while 3 submitted $\delta_{pred}$ and $\sigma_{pred}$ in addition. We exclude them from the following evaluations.  }
\label{fig:participant_flow}
\end{figure}
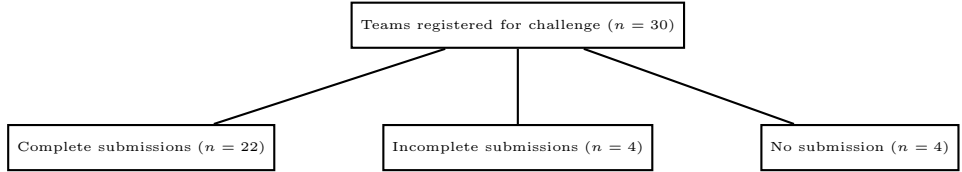    

\begin{table}[ht]
\caption{Final challenge ranking by evaluation criteria. The table also shows the statistical methods used by the teams (primarily based on their short methods description) and size of the Subgroup ($N$). The column names stand for variable selection (var. sel.), penalized regression (pen. regress.), tree-based methods (tree), forest/boosting (forest/boost.), and cross-validation (cross-val.). }
\label{tab:final-results}   
\centering
   \definecolor{r22}{gray}{0.440}
   \definecolor{r21}{gray}{0.462}
   \definecolor{r20}{gray}{0.484}
   \definecolor{r19}{gray}{0.506}
   \definecolor{r18}{gray}{0.528}
   \definecolor{r17}{gray}{0.550}
   \definecolor{r16}{gray}{0.571}
   \definecolor{r15}{gray}{0.593}   
   \definecolor{r14}{gray}{0.615}
   \definecolor{r13}{gray}{0.637}
   \definecolor{r12}{gray}{0.659}
   \definecolor{r11}{gray}{0.681}
   \definecolor{r10}{gray}{0.703}
   \definecolor{r9}{gray}{0.725}
   \definecolor{r8}{gray}{0.747}
   \definecolor{r7}{gray}{0.769}   
   \definecolor{r6}{gray}{0.790}
   \definecolor{r5}{gray}{0.812}
   \definecolor{r4}{gray}{0.834}
   \definecolor{r3}{gray}{0.856}
   \definecolor{r2}{gray}{0.878}
   \definecolor{r1}{gray}{0.900}    
   \definecolor{blk}{gray}{0.95}       
    \centering
    \begin{tabular}{ccccccccccccc}
\toprule
Team &  var. sel. & pen. regress. & tree & forest/boost. &  other & cross-val. & bootstrap & $N$ & $Z_1$ & $Z_2$ & $Z$  \\ 
\midrule
1    &    x       &     x         &   x  &               &        &    x       &           & 106 &  1.50 & 1.13  & 1.31 \\ 
2    &    x       &     x         &   x  &      x        &        &    x       &    x      & 107 &  1.61 & 0.94  & 1.27 \\ 
3    &            &               &      &               &   x    &            &           & 205 & 1.83  &-0.03  & 0.90 \\ 
4    &            &     x         &      &               &        &            &    x      & 213 & 0.66  &0.80   & 0.73 \\ 
5    &            &     x         &      &      x        &   x    &    x       &    x      & 73  & 1.35  &0.01   &0.68  \\ 
6    &    x       &     x         &   x  &               &        &    x       &           & 160 & 0.67  & 0.53  & 0.60 \\ 
7    &            &     x         &   x  &      x        &        &    x       &           & 306 & 0.93  &-0.01  & 0.46 \\ 
8    &    x       &     x         &   x  &               &        &    x       &    x      & 68  & 1.51  &-0.59  & 0.46 \\ 
9    &    x       &               &      &               &   x    &            &           & 182 & 0.28  &0.59   & 0.44\\ 
10   &    x       &     x         &   x  &      x        &   x    &    x       &    x      & 187 & 0.06  & 0.65  & 0.35 \\ 
11   &    x       &               &      &               &   x    &            &           & 143 &-0.08  & 0.66  & 0.29 \\      
12   &            &               &   x  &      x        &        &    x       &           & 64  & 0.45  &-0.12  & 0.16 \\ 
13   &            &     x         &      &      x        &   x    &    x       &           & 195 &-0.25  & 0.34  & 0.05 \\ 
14   &    x       &     x         &   x  &               &        &            &    x      & 84  &-0.79  & 0.34  & -0.22\\ 
15   &            &     x         &   x  &      x        &        &    x       &    x      & 134 &-1.45  & 0.81  & -0.32\\ 
16   &    x       &               &   x  &      x        &        &    x       &           & 199 &-0.06  & -0.68 & -0.37\\ 
17   &            &               &   x  &               &        &    x       &    x      & 57  & -0.07 & -3.63 & -1.85\\ 
18   &            &               &   x  &               &   x    &            &           & 60  & -0.57 & -3.54 & -2.06\\ 
19   &    x       &     x         &   x  &      x        &   x    &            &    x      & 251 & -1.38 & -3.17 & -2.27\\ 
20   &            &               &   x  &      x        &        &    x       &    x      & 83  & -1.99 & -10.02& -6.00\\ 
21   &    x       &               &   x  &               &        &            &           & 102 & -0.68 & -11.81& -6.25\\ 
22   &            &     x         &   x  &               &        &    x       &           & 68  & -0.33 & -17.16& -8.75\\ 
\bottomrule
    \end{tabular}
 
\end{table}

As described in Figure~\ref{fig:participant_flow}, from the 30 registered teams, 26 submissions were received, 22 of which we were valid in the sense of a completely populated submission file. 

Table \ref{tab:final-results} displays the final team submissions ranked by the primary evaluation metric $Z$, also included are $Z_1$, $Z_2$ and the size of the proposed subgroup ($N$) on the INVIGORATE 2 data. In Table \ref{tab:final-results} we provide a short summary of the utilized analytical strategy by the different teams as judged from the organizing team by their short methodological description and selectively inspecting the team's underlying programming code in case the description was not clear. A model is denoted as being used even if the team used this as part of their model selection strategy but did not select it for their final predictions. The column "var. sel." assesses whether the team did a preliminary variable selection before starting the modelling (either data-based or based on substantive background knowledge). The column "pen. regress." assesses whether a team used penalized regression (for example LASSO). The column "tree" assesses whether the team used a single tree approach, for example model based partitioning (MOB) \citep{zeileis2008model}. The column "forest/boost." assesses whether random forest or boosting techniques were utilized. And finally the "other" column includes other approaches that the teams used, like standard regression techniques, generalized additive models or other nonparametric/nonlinear regression methods, association rule learning and visual inspection of plots. The columns "cross-val." and "bootstrap" assess whether the teams used cross-validation or bootstrap in any form.

We describe the strategy of the top 5 teams in more detail
\begin{itemize}
    \item Team 1 pre-processed the data with missing data imputation (variables with more than 25\% missing values were removed), skewed data transformation, non-normal data normalization. Then the team compared the performance of virtual twins\citep{foster2011subgroup} based on LASSO, the model-based partitioning (MOB)\citep{zeileis2008model} using the train and test data split from the bootstrapped data, where the final model was selected based on the majority vote (1000 times). With the MOB method selected, the final subgroup was decided accordingly.
    \item Team 2 performed a covariate pre-selection based on discussion with one of the consulting clinicians and literature. Based on this the team compared MOB, LASSO (directly performed on subgroup variables with cut-offs for continuous variables based on medians) and causal forest \citep{wager2018estimation} using leave-one-study-out cross-validation and decided to submit the subgroup based on the results from the LASSO approach.
    \item Team 3 used the `pysubgroup` Python package \citep{lemmerich2018pysubgroup}, which implements association rule learning methods for subgroup selection, combined with cross-validation methods.
    \item Team 4 used Bayesian dose-response modelling based on the sigmoid Emax model to determine predictive variables \citep{thomas2022identifying}. Then a bootstrap approach for cut-off selection for the identified variables was used to come up with a subgroup.
    \item Team 5 selected 3 studies as the training data and 1 study as test data.  To begin with, the team took into account what was known scientifically of the disease (comedications, confounders, risk-benefit) as well as the co-medications. After some data wrangling (e.g., examining covariate colinearities and missing values patterns), the team used the SuperLearner\citep{van2007super} approach to predict treatment differences for all patients using relevant variables, and a few machine learning models. In addition of the built-in cross-validation of SuperLearner, they cross-validated the whole process of estimation. Based on these treatment differences, predictive variables were then selected using the knockoff algorithm. Finally, a Generalized Additive Model helped identify optimal cut-off points for continuous variables and to generate candidate subgroups in a multivariable model. Among candidate subgroups, the one with the highest score on the test data was submitted and bootstrap was used to estimate standard errors.
\end{itemize}

From Table \ref{tab:final-results} or the strategy from the top 5 teams no clear pattern emerges of whether a particular ingredient to a team's workflow lead to better results.

The task provided many options and decisions for the teams to make: Should a covariate pre-selection be made, before starting modelling? If yes how to do the pre-selection? Based on initial data analyses (not using the outcome variable)~\cite{Baillie:IDA:2022}, simple univariate analyses already using the outcome, discussion with the clinical experts, incorporating the scientific literature? At the modelling stage different modelling strategies can be utilized, but models for the outcome or the treatment effect don't directly provide "subgroups", which means that many teams did extra steps either at the input to these models (e.g. LASSO on subgroup indicators) or post-processing the output (e.g. using virtual twins based on LASSO). So even if teams used a particular method like LASSO this could be done in many different ways. Finally the teams implemented bootstrap or cross-validation in quite different ways to assess the quality of the subgroup proposed by the analytical workflow. For example some teams performed cross-validation on the study level, others just randomly left out a part of the data of the size of the INVIGORATE 2 trial.

We believe that the nature of subgroup identification (not readily available from model outputs; no fully established consensus on workflow) increased the variability in the employed analytical approaches. The variation in decisions were seen along the analytical pipeline \cite{Leek2015-id}, not just the selection of statistical model. Even if teams may have used the same or similar methods, they may have used different variants of the same method or combined different methods in different ways, ending up with a different subgroup found. Therefore, it is challenging to draw strong conclusions on whether particular "methods" performed better than others based on Table \ref{tab:final-results}. The problem definition left teams with the freedom to specify an analysis strategy that went beyond specifying a model only. This is an illustration that the flexibility of exploratory questions provides many choices. The exercise revealed the many (often hidden) choices available when defining the analytical strategy and executing this strategy. 

\subsection{Comparison of submitted subgroups}

\begin{figure}
    \includegraphics[width=\textwidth]{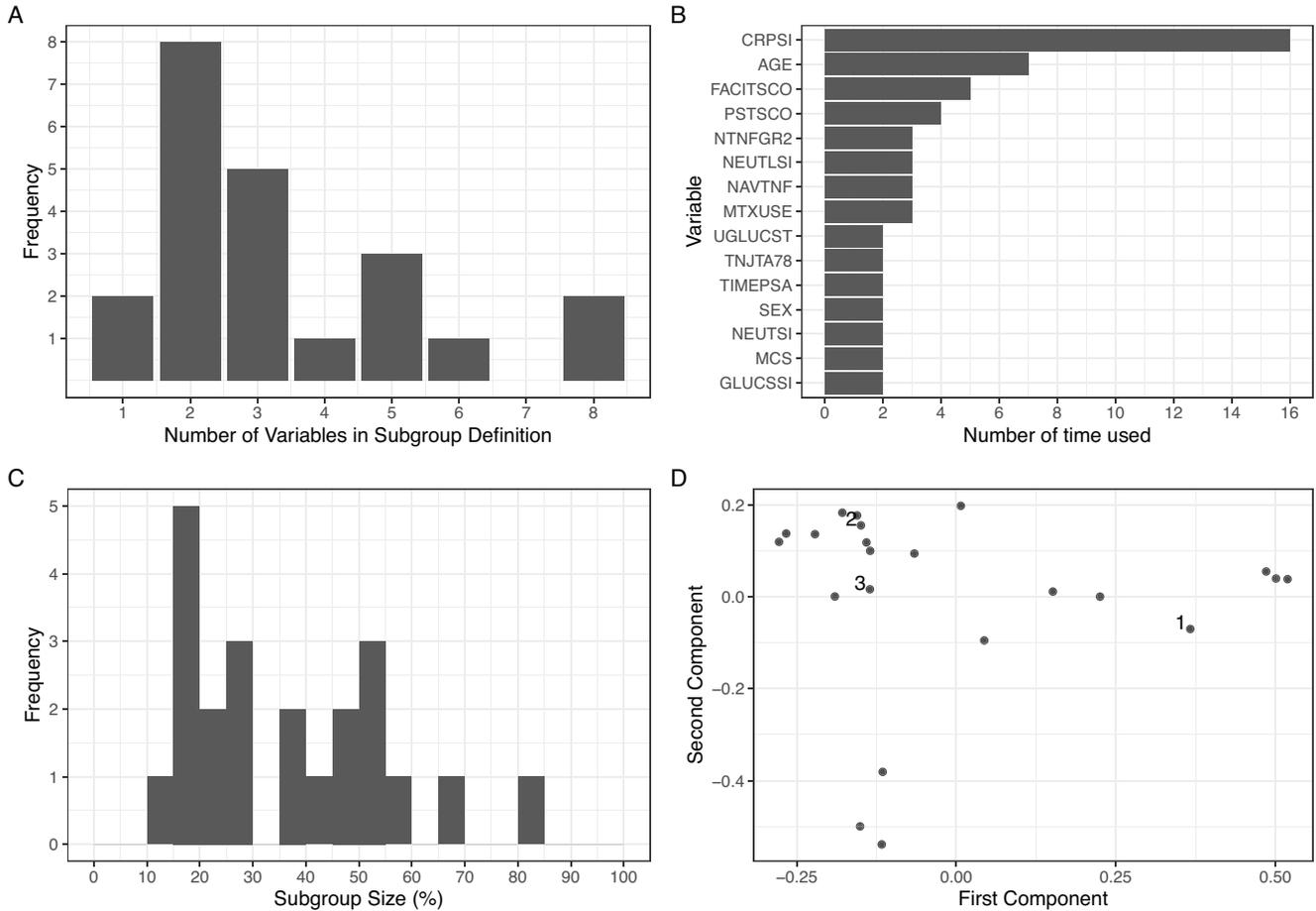}
    \caption{Characteristics of the subgroups submitted by the participants. A) Number of variables used in the subgroup definition; B) Frequency of appearance of each variable subgroup definitions (see Table \ref{tab:var_abbrev} for the variables underlying the acronyms). Only variables used at least twice are included in the plot; C) Histogram of subgroup sizes, measured as the percentage of all patients that belong to the subgroup;  D) First two principal components computed based on pairwise Jaccard distance between the subgroups, the top 3 teams are marked by the corresponding number.}
    \label{fig:res1}
\end{figure}

Subgroups submitted by the different teams also differed considerably (see Figure \ref{fig:res1}) in terms of which and how many variables were used for the subgroup definition (Figure \ref{fig:res1}A), as well as in subgroup size (Figure \ref{fig:res1}C), ranging between $10$ and $85\%$ of the trial patients. 

The most used variables were c-reactive protein (CRP), a measure of inflammation, patient age in years at randomisation (AGE), and a measure of patient fatigue (FACIT score)  (see Figure \ref{fig:res1}B). The emergence of these variables on the aggregate level across teams is in agreement with a previous analysis of the FUTURE 2-5 data \citep{sechidis2021using}, where a clinical interpretation of these results can be found. 

We also investigated the similarity of the submitted solutions by computing the pairwise Jaccard distance of the subgroups. The Jaccard distance is defined by 1 minus the cardinality of the intersection of the two compared subgroups divided by the cardinality of the union of the two subgroups. Multidimensional scaling was performed and the first two principal coordinates are displayed (Figure \ref{fig:res1}D). 

Interestingly, the subgroup proposed by the top team (highest $Z$) is quite different from the ones proposed by the 2nd and 3rd team, if compared to the overall diversity of the proposed subgroups.

\subsection{Treatment effects of submitted subgroups}

\begin{figure}
\includegraphics[width=\textwidth]{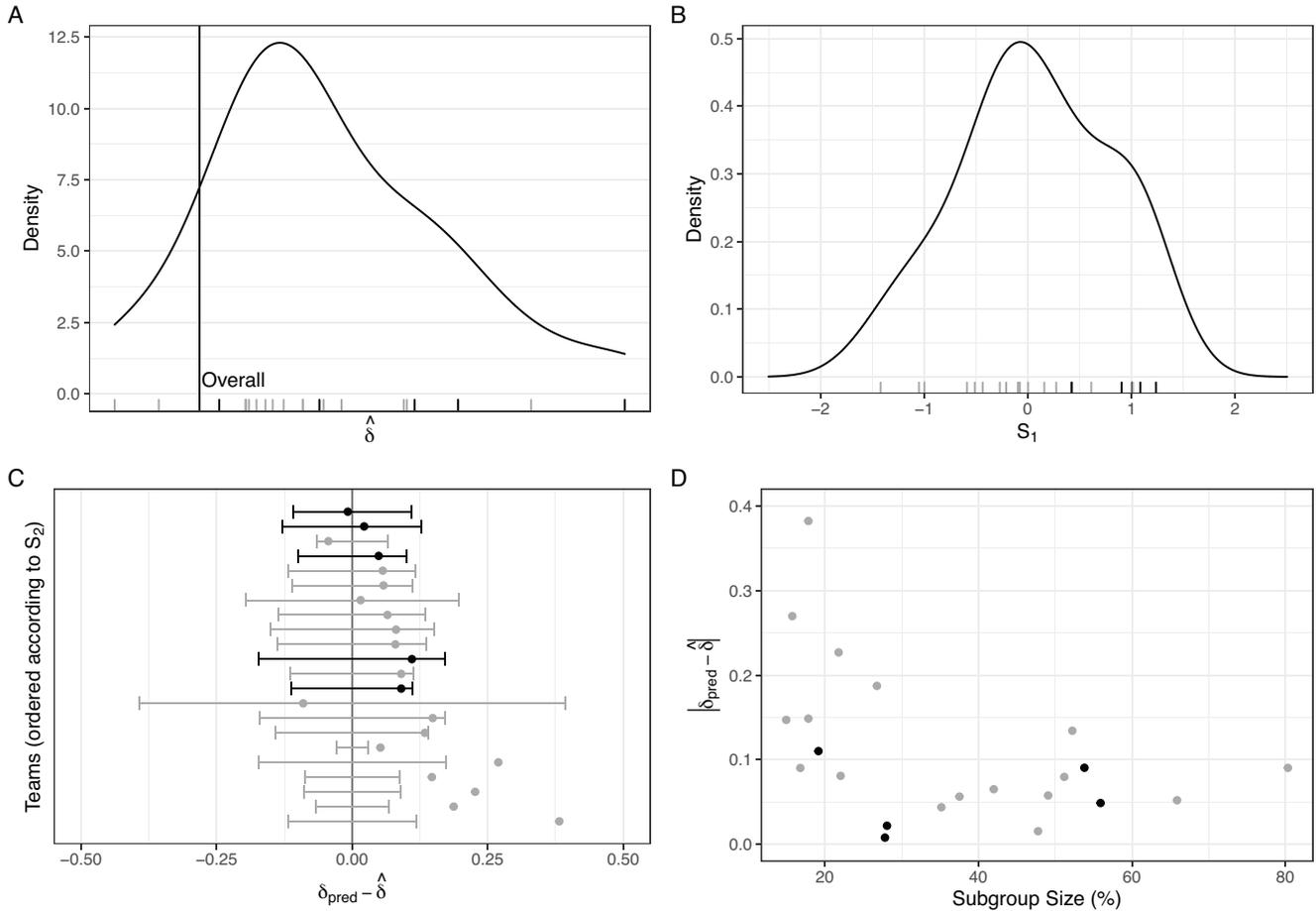}
\caption{A) Distribution of the treatment effect $\hat{\delta}$ observed in the proposed subgroups in INVIGORATE 2. The vertical line marks the overall treatment effect in the trial. Ticks on the x axis mark the results of single teams, black ticks represent top 5 teams; B) Distribution of the score $S_1$. Ticks on the x axis mark the results of single teams, black ticks represent top 5 teams; C) Difference between the treatment effect predicted by each team for their subgroup, $\delta_{pred}$, and that observed for that subgroup in INVIGORATE 2, $\hat{\delta}$. The $95\%$ CI based on $\sigma_{pred}$ is also shown. Teams are ordered according to their $S_2$ score, the top 5 teams are in black; D) Modulus of prediction error plotted against subgroup size, each dot represents a team, top 5 teams are in black.}
\label{fig:res2}
\end{figure}

The treatment effect of the submitted subgroups on risk difference scale in the INVIGORATE 2 trial was, in all except two cases, higher than the overall treatment effect (Figure \ref{fig:res2}A). Interestingly, the distribution of the first score $S_1$, which would be centered around zero if the subgroups were chosen randomly, is only slightly skewed towards positive values (Figure \ref{fig:res2}B). The reason for this is that $S_1$ measures the interaction on the logit scale, and in this case the two scales (risk difference vs logit scale) don't lead to entirely consistent results. We attribute this to the non-linearity of the logit scale. An increase of the probability of response under treatment causes only a small, sub-linear increase of $S_1$, while decrease in the probability in response on placebo is rewarded much higher. It is well-known that the concept of interaction depends on the scale of the effect measurement so this result is not surprising. In fact there are empirical observations as well as mathematical results indicating that the logit scale leads to a more homogeneous effect measure compared to the risk difference scale \citep{poole2015risk, ding2015differential}, which may also explain the results here. See Appendix \ref{app:scores} for more details on the differences between the risk difference and logit scales and for a sensitivity analysis on how the ranking would change if we used a score on the risk difference scale. 

The prediction that the participants provided for the treatment effect in their subgroup based on the FUTURE 2-5 studies, $\delta_{pred}$, was almost always larger than the actual treatment effect observed for the same subgroup in the INVIGORATE 2 trial, $\hat{\delta}$ (Figure \ref{fig:res2}C). The average percentage overestimation across teams was $34\%$. We attribute this observation to the `regression to the mean' effect. The $95\%$ prediction intervals based on the submitted standard deviations $\sigma_{pred}$ mostly cover the observed value but for less teams than expected (Figure \ref{fig:res2}C). 

The prediction error is larger, in magnitude, for small subgroups (Figure \ref{fig:res2}D). This trend may be due to a larger "regression to the mean" effect in small samples, where the variability is larger. From Figure \ref{fig:res2}D it is also interesting to see that the overall top 5 teams proposed subgroups of quite different sizes (black dots), with one as small as $20\%$ and other teams with subgroups of around $58\%$.
 
The $Z_1$ and $Z_2$ scores have some positive correlation at low values, i.e. teams with a poor performance in one score tended to have a poor performance also in the other, but this correlation disappears at higher values (see Figure \ref{fig:scores}A in the Appendix). The two scores are similarly correlated with the final score $Z$ (Spearman correlation, respectively, 0.87 and 0.74), such that the final team ranking is roughly equally influenced by each. Indeed, we can observe in Figures \ref{fig:res2}B and C that the top 5 teams according to $Z$ (highlighted in black) also score high according to $S_1$ and $S_2$. 

To assess robustness of the ranking to the score definition, we tested an alternative method to rank the teams, where we first compute two separate ranking, one according to $Z_1$ and one according to $Z_2$, and then we average the ranks to obtain the final score. This ranking method produces a similar top 5 grading.

\subsection{Feedback Survey: the participants view}
After communication of the results, a feedback survey was sent to the participants and filled out by 65 participants. The survey had 17 questions, including multiple-choice questions to rate satisfaction and organizational aspect and questions allowing for free-text answers to collect more general feedback and what participants learned and recommend regarding subgroup identification. 

Overall, the challenge was very well perceived by the participants. On a scale from 1 (bad) - 5 (great) for the question "How do you rate your overall experience participating in the subgroup challenge?", $82\%$ of the responses were 4 or 5, and $42\%$ were 5. We asked what people liked about the challenge as a free text question. A common feedback was that they enjoyed learning about a new topic, getting hands-on experience with the methodology, and appreciated the introductory material provided. Participants also judged positively the realistic setting of the challenge: the opportunity to test methods on real clinical data and the open-ended problem with unknown answer, like in practical work.  As part of the challenge participants worked on a company-internal new data science platform, and for many of them it was the first opportunity to try it. They enjoyed learning to use this platform and exploring its functionalities. Some participants mentioned that they enjoyed that the competition was global including all major sites of the company, and that it was an occasion to connect with colleagues beyond their usual circle. They enjoyed  working on something new and different compared to the normal work, in a competitive but friendly environment.

We also asked what people learned. Here individual participants mentioned that they enjoyed learning about subgroup identification methods (e.g. reading papers, exploring R packages). Additionally, they appreciated the occasion to improve their coding skills, discover new packages, and become more familiar with good data science practices that facilitate collaborative work, e.g. Git for version control.  With respect to learnings on subgroup identification, participants highlighted the learning that domain knowledge is important for this task and how challenging prediction of differential treatment effect is. Finally participants enjoyed learning more about the PsA disease area.

\section{Discussion and conclusions}
\label{sec:conc}

We conducted a company-internal data challenge around the topic of subgroup identification. Participants worked on available clinical trial data to propose a subgroup for an upcoming study.

Overall almost all teams succeeded in finding a subgroup with treatment effect estimate increased versus the overall treatment effect in the new study on the risk difference scale. However most of the teams overestimated the treatment effect of their subgroup in the new study, which we attribute to a regression to the mean effect.

There are several learnings emerging from this data challenge. The first and most important learning, also reflected in the responses to the feedback survey, is that the identification of subgroups with increased or decreased treatment effect is "difficult". Even in the ideal situation of the challenge, where four clinical trials have been available for learning the subgroup definition, finding subgroup results that can be reproduced in a single new trial turned out difficult. Almost all teams were able to find subgroups with treatment effects larger than the observed overall treatment effect, but the treatment effect prediction for the subgroup was optimistic for most teams. The quality of the prediction depended on the size of the subgroup (small subgroups are more challenging to predict). This observation is in-line with pharmaceutical statistics "folklore" and we believe this is an important practical experience. To enable this we conducted the challenge in a "safe space" for exploration and testing, outside of the project related work, while still being as realistic as possible. In addition, results of the teams were presented in a semi-blinded fashion: Only the team names of the top 5 teams were revealed publicly, so that participants were not concerned about potential negative consequences of performing poorly in the challenge (which could have led to not participating or submitting a result). 

With respect to subgroup identification, there were no concrete learnings in the sense that particular analytical methods worked better than other methods. The reason is that for subgroup identification several steps are required to come up with a subgroup, as most analytical approaches for modelling do not directly report a subgroup. In addition, most teams combined analytical approaches with meta-methods such as bias adjustment or cross-validation techniques, effectively making a clear comparison difficult. The breadth of the provided solutions illustrated that company internal guidance on the topic may be needed, to avoid that teams "re-invent the wheel". In a follow-up session to discuss the challenge results, participants also raised the question, whether the focus should really be on "subgroups" with increased or decreased treatment effect, rather than focusing on identifying covariates that may modify the treatment effect. The additional step towards defining concrete subgroups may not be needed. In addition reporting selected subgroup results only, may not provide a good overview of the treatment effect heterogeneity observed in the trial.

With respect to the challenge format, our anticipation was confirmed (at least according to participant feedback) that such a challenge is an effective way of learning on an individual level, about subgroup identification, about software tools such as R, Rmarkdown, Git and the new scientific computing platform as well as the disease indication. Participants also mentioned the benefit of learning within the team from other, more experienced, team-members. After the event informal and also formal (as part of the result discussions) learning continued also across teams, and experiences and learnings from the top 5 teams were shared: The CTF, i.e. common task, data and rules provides a "lingua franca" between participants to discuss the problem and analytical strategies \citep{Baker2012-as}.

What are the weaknesses of the challenge as it was conducted? We decided to utilize a real case example with real unknown results on the new study. This has the downside that we cannot tell for sure whether the "winning" subgroups are actually the best. The new study is just "one further study" and when looking into subgroups in one study, for the purpose of scoring, there will be randomness in the subgroup results. This was also communicated as a limitation to the participants at the kick-off meeting. 

In this challenge all participants had to submit a subgroup. Before the challenge we considered alternative rules where teams would not need to provide a subgroup, but have the option to submit the overall trial population and a prediction for the overall treatment effect. We decided against this option, to keep the challenge rules simple, but in retrospect this still sounds like an attractive option to reconsider.

An alternative to a real case example would be to run a simulation challenge, where the organizing team can control the truth. In such a challenge participants could submit algorithms and the repeated sampling performance could be assessed against the real simulation truth. In such a settings teams would have been forced to program a clear strategy for subgroup identification (which would also have made assessment of solutions easier). The downside of this approach is that the task of the participants would then essentially be to guess the simulation truth that the organizers consider plausible. Even if multiple scenarios are utilized this remains an important disadvantage from our perspective.

In this challenge we used a real "hold-out" study with primary results not yet published and under restricted access. At the end of the challenge several participants showed interest in exploring the data from the hold-out study themselves. Unfortunately this was not possible as the access restrictions on the regulatory data platform were still in place. In addition no anonymized data set (that could be shared with participants) was available. While this is a downside of the utilized challenge setup, we believe the benefit of using a real future study with unknown results for scoring (and thus mimicking reality better) outweigh this disadvantage.

Finally we provided introductory material and code for participants to get started. Based on the feedback this was appreciated by the participants to get started into the topic methodologically, while directly having code available. But there is a trade-off to consider: Many teams utilized the provided methods as a starting point for their own investigations and thus had an influence on the participants solutions. For example we were surprised by how few teams utilized traditional standard statistical methods for the purpose of this challenge (e.g. regression based methods).

\textbf{ACKNOWLEDGEMENTS}\\
We like to thank all participants of the challenge, and would like to particularly highlight the members of the top 5 teams Team 1 (Team "GO SUBGROUP GO!") Mingmei Tian, Tao Li, Yongmin Liu, Pengpeng Wang; Team 2 (Team "Better Than Placebo") Beilei He, Markus Lange, Jelena Čuklina, Abdelkader Seroutou; Team 3 (Team "AI4Lifers") Joana Marques Barros, Matthias Hueser, Hagen Muenkler; Team 4 (Team "Rgonauts") Isidoros Papaioannou, Marius Thomas, Zheng Li and Team 5 (Team "The Counterfactual Knockoffs") Sandra Lopez-Leon, Rima Izem, Matthias Kormaksson. We also would like to thank the two clinicians who were available for consulting: Aimee Readie and Sarah Whelan. We would like to thank the whole Cosentyx team for the support of this challenge, in particular Xuan Zhu and Chengeng Tian. Furthermore we would like to thank: Brian Porter, Janice Branson and Eric Gibson, who were part of the qualitative reviewers of the methodological statements; Kostas Sechidis and Frank Bretz, who helped with the testing and review of the material (Git template repository, challenge rule-book) provided to the participants; Jana Stárková, who supported as part of the organisation team towards the end of the challenge and finally Björn Holzhauer for ideas and discussions around the design and scoring of the challenge and proposing the idea of using a "real hold-out" (i.e. still access-restricted) study for scoring.

\bibliographystyle{abbrv}

\appendix

\section{Difference between optimizing the treatment effect on the risk difference and logit scales}
\label{app:scores}

The score $S_1$ measures the treatment effect on the logit scale. While on the linear scale an increase of the response under treatment and an identical decrease of the response under placebo would affect the treatment effect in the same way, the treatment effect on the logit scale is more sensitive to the response under placebo than to the response under treatment. 

\begin{figure}
\centering
\includegraphics[width=\textwidth]{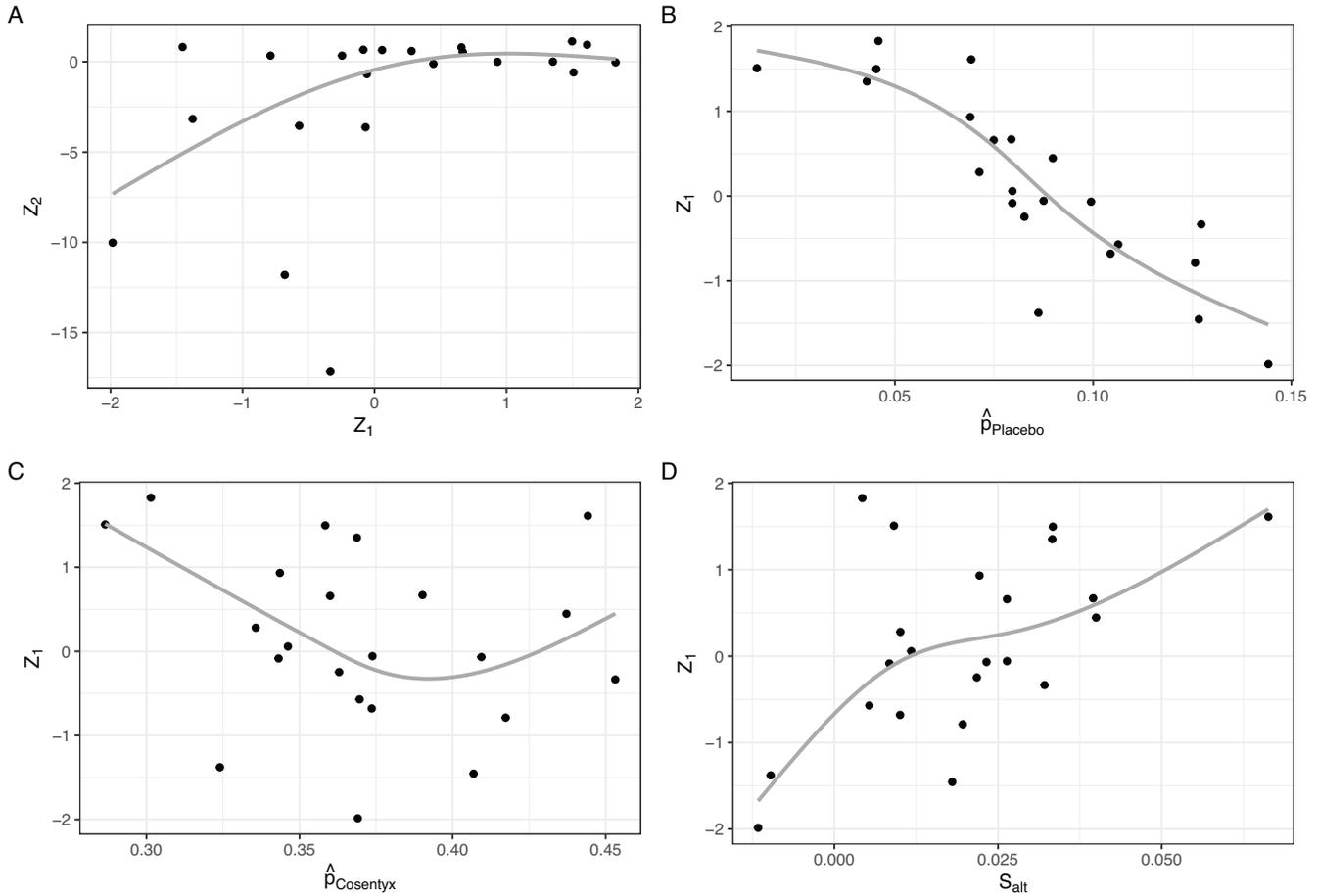}
\caption{A) Scatterplot of the two scores $Z_1$ and $Z_2$; B) Score $Z_1$ plotted against the probability to respond on placebo in the subgroup; C) Score $Z_1$ plotted against the probability to respond on treatment in the subgroup;  D) Score $Z_1$ plotted against the alternative score on the risk difference scale,  $S_{alt}$ for $\alpha=0.5$. In all panels, each dot represents a team and the grey smoothing lines are cubic regression splines with 3 degrees of freedom.}
\label{fig:scores}
\end{figure}

Indeed, $S_1$ is anti-correlated with the probability of response under placebo (see Figure \ref{fig:scores}B), while it has a weaker dependency on the probability of response under treatment (see Figure \ref{fig:scores}C). This can be explained with the non-linearity of the logit function, which grows faster at smaller values, and implies that minimizing the response under placebo in the subgroup is a more effective strategy to get a high $S_1$ score, compared to maximizing the response under treatment. 

To test the robustness of the ranking with respect to the choice of measuring the treatment effect on the risk difference or on the logit scale, we performed a sensitivity analysis where we compare the original ranking to one obtained with an alternative definition of the $S_1$ score on the risk difference scale. We defined $S_{alt}= f_{sub}^\alpha (\hat{\delta}_{sub}-\hat{\delta}_{overall})$, where $f_{sub}$ is the proportion of patients in the subgroup divided by the overall sample size, $\hat{\delta}_{sub}$ and $\hat{\delta}_{overall}$ are, respectively, the treatment effect observed in the subgroup and in all patients, and $\alpha$ is a parameter. The term $f_{sub}^\alpha$ is introduced to avoid rewarding very small subgroups that achieve a high treatment effect due to chance.  For $\alpha=0, 0.5$ and $1$, rankings slightly changed but were overall similar. This can also be seen from Figure \ref{fig:scores}D: The original $S_1$ score and the alternative one (for $\alpha=0.5$) are correlated overall, but there are some outliers. 

\begin{table}[ht]
    \caption{Variable description for the variables given in Figure \ref{fig:res1}B}
        \label{tab:var_abbrev}    
\centering
    \begin{tabular}{ll}
\toprule
Acronym &  Variable description \\ 
\midrule
CRPSI    &  C-reactive protein (mg/L) \\ 
AGE      &  Age (years)       \\ 
FACITSCO & Fatigue Total Score  \\ 
PSTSCO   &  PASI Total Score   \\ 
NTNFGR2  &  Number of prior anti-TNF   \\ 
NEUTLSI  &  Neutrophils/Leukocytes (\%)  \\ 
NAVTNF   & Naive to TNF Alpha Inhibitors\\ 
MTXUSE   & Methotrexate Use at Baseline \\ 
UGLUCST  & Urine Glucose Dipstick \\ 
TNJTA78  & Baseline Adjusted Tender Joint Total Score for PsA 78 Joints  \\ 
TIMEPSA  & Time since First PsA Diagnosis (Years) \\      
SEX      & Sex          \\ 
NEUTSI   &  Neutrophils (Absolute) (10E9/L)  \\ 
MCS      &  SF36-MCS   \\ 
GLUCSSI  &  Glucose, Serum, Fasting (mmol/L)  \\ 
\bottomrule
    \end{tabular}
\end{table}
\end{document}